# The CL and PL characteristic of different scale CsI:Na crystals

HU Jun-Peng(胡俊鹏)[1]   LIU Fang(刘芳)[1,2]   OUYANG Xiao-Ping(欧阳晓平)[1,2,3]

[1] School of Nuclear Science and Engineering, North China Electric Power University, Beijing 102206, China;

[2] Radiation Detection Research Center, Northwest Institute of Nuclear Technology, Xian 710024, China;

[3] Department of Engineering Physics, Tsinghua University, Beijing 100084, China;

**Abstract:** We investigate the luminescence characteristic of different scale CsI:Na crystals excited via photoluminescence(PL) and cathodoluminescence (CL). The CsI:Na crystals are processed to three samples with different diameter, decreasing from micro-scale to nano-scale. It is found that the nano-scale CsI:Na crystal emits 420nm luminescence by PL , while its emission band is at 315nm and 605nm by CL. The reason for this phenomenon relates to the energy and density of incident particles, and the diameter of CsI:Na crystal. When crystal diameter decreases to nano-scale, the number of surface defects relatively increases, leading to the Na-relative luminescence decreasing. In addition, we have also investigated the CsI:Tl crystal with the same experiment condition. The result indicates that the emission almost has nothing to do with the crystal diameter.

## 1 Introduction

Sodium-doped cesium iodide crystals have been widely used, as a very effective scintillation material, attribute to the high stopping power for X-rays and γ-rays from the relatively high density and large atomic numbers of both Cs and I[1]. The blue luminescence was first identified as due to the existence of the non-isomorphic impurity Na in CsI by Brinckmann [2]. In recent years, a lot of work in relation to luminescence mechanism of CsI:Na have been researched, and achieved certain results. It is found for the first time that the characteristic emission consists of two band, centered at 4200 Å and 3700Å, respectively, and the latter one increased as the temperature decreased[10]. The 2.95eV band of CsI:Na has been researched, which attributed to tunnelling recombination between Na-band electron and $V_k$ centers. When CsI:Na crystals diameter decreases to nano-scale, the X-ray excited luminescence decay time is speeding up from ~600ns to ~10ns [3]. In addition, recent studies have indicated that some of the emission bands are of excitonic origin are similar to those of alkali halides with the NaCl lattice [4,5]. The light output reaches a maximum at 0.01 mol. % sodium and decreases at higher sodium concentrations[6,7]. Nanophosphors have been extensively investigated due to their application potential for various high-performance[14].

At the same time, emission spectrum of CsI:Na usually is not well matched with the spectral sensitivity of photomultipliers, and the luminescence decay time is about 650ns, compared with 10ns[8] of pure CsI. All of these two shortcoming seriously restrict application CsI:Na crystal. But today the issue of luminescence decay time have been well resolved [3].

In the present experiments, we have investigated luminescence spectrum of CsI:Na with different diameter, which is excited via photoluminescence(PL) and cathodeluminescence(CL). It found that the emission spectra of different scale (range from centimeter scale to nano-scale) CsI:Na crystals are quite clearly different excited by CL compared with being excited by PL. Particularly, the luminescence peaks of the nano-scale CsI:Na crystal is at 315nm and 605nm excited by CL, however the emission band maximum is at 420nm, when excited by PL. Also, we have researched diverse crystal diameter of CsI:Tl with the same excited method, and the result indicated the spectrum nearly consistent excited whether by CL or PL.

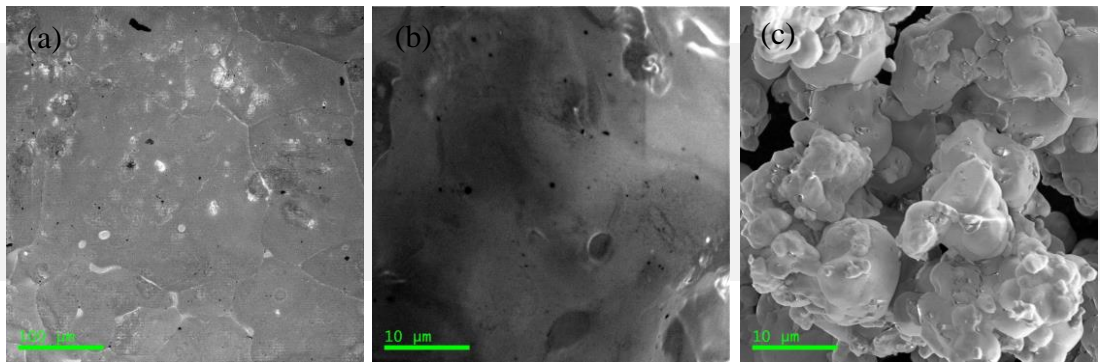

Fig. 1.  SEM images of CsI:Na crystals A,B and C .

## 2  Experimental methods

Firstly, samples of CsI:Na crystals are grown by means of Bridgman-Stockbarger method, and the doping concentration of $Na^+$-ions is about 0.02mol.%. The high resolution scanning electron microscope (Hitachi S-4800, Japan) is used to reveal the surface features and illustrate different diameter crystal samples. Fig. 1 is the the scanning electron microscope (SEM) images of different crystals diameter, and the different crystals are labeled as crystal A, crystal B and crystal C, respectively, whose diameter are millimeter scale, 20-50 μm and 200 nm-20 μm, respectively.

And then, in order to measure the luminescence spectrum of CsI:Na crystals, we make use of Cathode Luminescence (CL) image (Gatan Monocl, USA) and the scanning electron microscope (FEI Quanta.400FEG). On the other hand, for the measurement of PL spectrum, we stimulate the CsI:Na crystal samples with ultraviolet light of 265nm which is generated from a xenon lamp.

## 3  Experimental results and discussion

Fig. 2(a)-(c) shows the cathodoluminescence (CL) spectra of CsI:Na crystals  respectively corresponding with crystals A, crystals B and crystals C of Fig. 1. Under the stimulating of ultraviolet light of 265nm, it seen that the emission spectra of crystals B is similar to that of crystals A, even though the emission intense at each emission peak is diverse from each other. It exhibits a strong blue emission at 420nm with a half-width of 120nm under ultraviolet excitation, which had been observed at both LHe temperature and at room temperature, ascribed by various authors to pure CsI due to the existence of the non-isomorphic impurity Na in CsI by Brinckmann[2]. But Panova [10] et al. observed that the excitation for the blue luminescence from CsI Na is not fundamental. The other luminescence peaks are at 315nm and 605nm, and the 315nm component is intrinsic to the self-trapped excitons of host crystal, and the 605nm component is attributed to impurity defects.

Compared with crystals A and crystals B, whose diameter are millimeter scale, 20-50 μm, The CL spectra of crystals C shows that the 420nm peak is very week, while the measuring condition is the same. The phenomenon also exist at the thin film samples[10], which thinks when the sodium concentration in the films is 0.01 mol. % the average distance between two sodium ions in a row, and the emission from centers is too week to be detected.

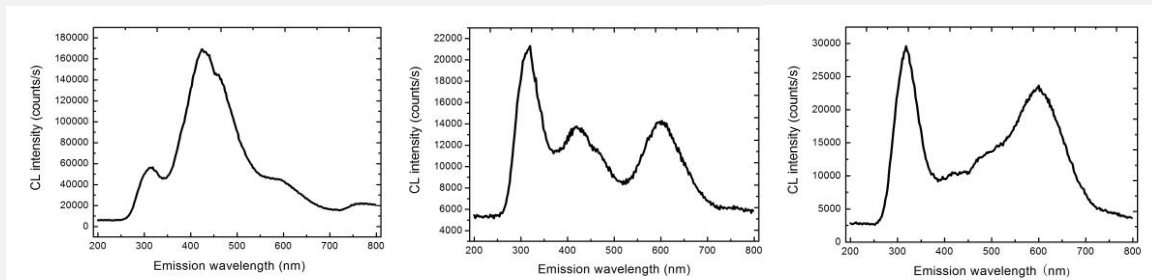

Fig. 2. CL emission spectrum of CsI:Na crystals A , B and C

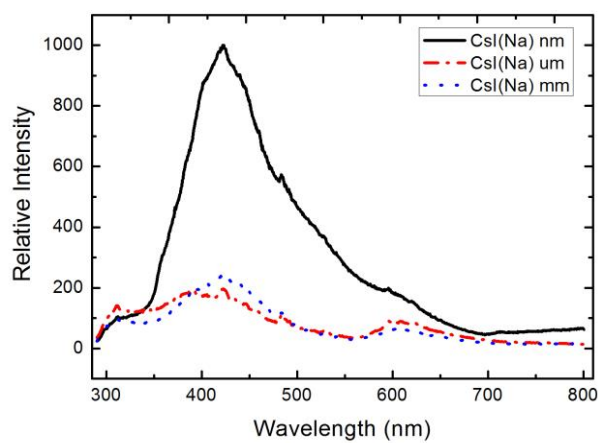

Fig. 3. PL emission spectrum of CsI:Na crystals A (blue), B (red) and C (black)

The reason for the phenomenon is explained clearly by Liu [3], thinks that CsI:Na crystal diameter impacts luminescence spectrum excited via CL, the CsI(Na) crystal diameter reduction leads effect of surface, With the diameter of crystal CsI:Na decreasing with excitation at 265nm. centimeter scale to about nano-scale, the Na-related luminescence becomes week. Hence, it occurred that emission peak at 420nm is weaken.

Next, we investigate the PL emission spectra of CsI:Na crystals excited c via UV light of 265nm. The emission spectra of crystals A, crystals B and crystals C are shown in Fig. 3. The emission peaks of crystal A and crystal B is almost coincident. The emission spectrum is consisted of two peaks centered at 420nm, 600nm-610nm and 315nm. Whereas, as for the emission spectrum of nano-scale CsI:Na crystal. it nearly only shows an intense single emission centered at 420nm.

Comparing with the emission spectrum peaks of CsI:Na crystals excited via CL, we can find out that The 420nm emission peak intensity nano-scale CsI:Na crystal is different via different activation manners (PL and CL), and these three CsI(Na) crystal samples are measured under the same work conditions. The PL and CL emission spectra of CsI:Na crystal samples are listed in table 1. It is noticeable that the emission spectra peaks of millimeter scale and micrometer scale crystals is almost no difference, but the experimental result is opposite, just as the Table 1 exhibitting. The emission peaks of crystal C excited by CL are at 315nm and 605nm, while at 420nm by PL, and the phenomenon have not been found so far.

We found many distinctions between above-mentioned activation manners. Energy of cathode rays is thousands of times than the UV. High-velocity electrons will ionize electrons in the crystal, and make them being of huge kinetic energy, which are so-called secondary electrons, and because the effect of surface, owning to the reduction of crystal diameter, the effect is enhanced with each other between initial electron these will generate the secondary electrons. The density of the secondary electrons is huge, the finally exciting the crystal to luminescence. In addition, the number of secondary electrons escape out of crystal surface decreases, that is another difference between the two exciting manners of CL and PL.

With the decreasing of CsI:Na crystal diameter, the number of surface defects relatively increases. When crystal C is excited by CL, the vast electrons resist the hole being trapped by Na +-ions. Whereas, recombination of self-trapped hole with Na0, arising from 420nm luminescence. But, when UV stimulates the crystal, due to the low density of UV and low energy of UV. it can be concluded that the main drawback of nano-scale phosphors is their lower quantum efficiency compared to microscale particles. This is attributed to the large surface area, which amplifies quenching processes[13].

Table 1  The emission spectrum of CsI:Na crystal samples excited by PL and C.

| CsI:Na crystals | Diameter | Spectrum peaks excited via CL/(nm) | Spectrum peaks excited via PL/ (nm) |
|---|---|---|---|
| A | 0.05(mm)-0.864(mm) | 315, 420, 460, 590 | 315, 420, 590 |
| B | 20(μm)-50(μm) | 315, 420, 605 | 315, 385 420, 480, 605 |
| C | 200(nm)-20(μm) | 315, 605 | 420 |

In addition, we also have investigated the emission spectra of different scale CsI:Tl crystals by PL and CL. The CsI:Tl crystal samples of different diameter are prepared by the same methods (Bri- dgman-Stockbarger).

The Fig. 4 shows the emission spectra of CsI:Tl crystals by PL. Like the above mentioned CsI:Na crystals, the CsI:TI crystals are also produced to being of different diameter, that is millimeter scale, micrometer scale and nanometer scale. From the Fig. 4 we found that the scintillation light of CsI(Tl) crystals is emitted in a wavelength band situated around 540 nm. There is no obvious changes at aspect of the luminescence spectra when the crystals diameter decreasing from millimeter scale to nanometer scale.

As arising from a progress in which the $Tl^+$ first captures an electron. When doping the Tl in the CsI, the STE radiation will be absorbed within the crystal, and this will contribute to generate $Tl^+$ luminescence[12]. However at the stimulating of cathode ray, the emission spectra of CsI:Tl crystals situate at 560 nm in Fig. 5, have nothing to do with the crystal diameter. We think that the reason of this phenomena is the different impurity in purty CsI has different luminescence mechanisms.

It would be of considerable interest to examine separately the intensities of the Tl band and the ultra-violet emission band which is characteristic of the pure material, as a function of ionization density and Tl content Unfortunately, this is not possible in NaI(T1) or CsI(T1), as a consequence of the overlap between the ultraviolet band and the Tl absorption[13]. In addition, the Tl+ ions emission dominates the CsI(Tl) cathode luminescence and light peaks have stable, relative intensity.

Submitted to 'Chinese Physics C'

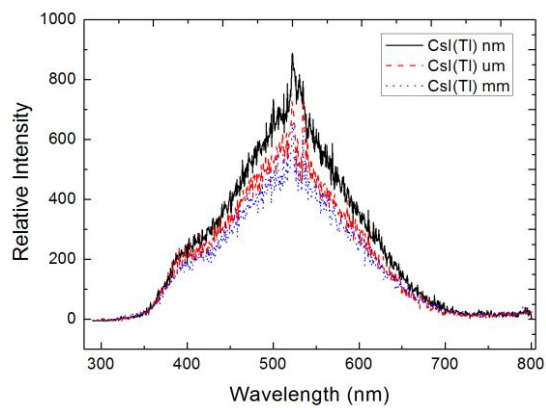

Fig. 4. PL emission spectrum of CsI:Na crystals A (blue), B (red) and C (black) with excitation at 265nm

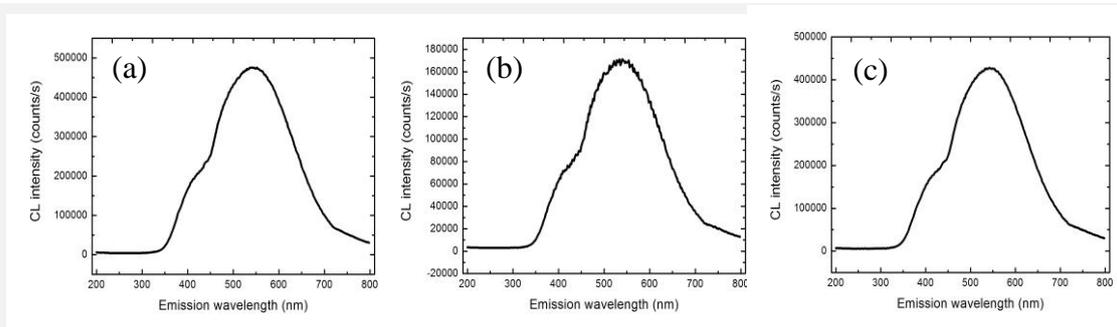

Fig. 5　CL emission spectrum of CsI:Tl crystals A , B and C

## 4　Conclusion

In our work, we have investigated the emission spectra of CsI:Na ranged from millimeter scale to nano-scale excited by CL and PL. The result indicates that the emission spectra of nano-scale CsI:Na crystals lose the peak at 420nm by cathode luminescence, which is due to the recombination of STEs and $Na^0$, while this phenomenon have not being when using ultraviolet light of 265nm to stimulate nano-scale CsI:Na crystals. The reason of this result is that the impacting of diameter decrease and different stimulation mechanism of PL and CL. In addition, the emission spectra of CsI:Tl crystals have also been briefly researched, and the result shows that the emission spectra have no obvious changes in spite of the crystals diameter decreasing to nano-scale due to $Tl^+$ ions emission dominating the CsI(Tl) luminescence.

## ACKNOWLEDGMENTS

This work is supported by Beijing Higher Eduction Young Elite Teacher Project (No. YETP0720).